\documentclass[12pt]{article}
\usepackage{graphicx}
\usepackage{xspace}
\usepackage{wrapfig}

\newcommand\pubnumber{SNSN-323-63}
\newcommand\pubdate{\today}

\def\institute{Czech Technical University in Prague, CZECH REPUBLIC}

\def\Title#1{\begin{center} {\Large #1 } \end{center}}
\def\Author#1{\begin{center}{ \sc #1} \end{center}}
\def\Address#1{\begin{center}{ \it #1} \end{center}}

\newcommand\pubblock{\rightline{\begin{tabular}{l} \pubnumber\\
         \pubdate  \end{tabular}}}
\newenvironment{Abstract}{\begin{quotation}  }{\end{quotation}}
\newenvironment{Presented}{\begin{quotation} \begin{center} 
             PRESENTED AT\end{center}\bigskip 
      \begin{center}\begin{large}}{\end{large}\end{center} \end{quotation}}

\def\beq{\begin{equation}}
\def\eeq#1{\label{#1}\end{equation}}
\def\eeqn{\end{equation}}

\def\beqa{\begin{eqnarray}}
\def\eeqa#1{\label{#1}\end{eqnarray}}
\def\eeqan{\end{eqnarray}}

\let\bar=\overbar

\def\etal{{\it et al.}}
\def\ie{{\it i.e.}}

\def\Dslash{\not{\hbox{\kern-4pt $D$}}}
\def\dslash{\not{\hbox{\kern-2pt $\del$}}}

\def\ttbar{\ensuremath{t\bar{t}}\xspace}
\def\ppbar{\ensuremath{p\bar{p}}\xspace}
\newcommand{\met}{\ensuremath{{}/\!\!\!{p}_{\rm T}}\xspace}
\newcommand {\pt}       {\ensuremath{p_T}}
\newcommand {\pythia}   {\textsc{pythia}\xspace}

\newcommand {\alpgen}   {\textsc{alpgen}\xspace}
\newcommand {\mcatnlo}  {\textsc{mc@nlo}\xspace}
\newcommand {\herwig}   {\textsc{herwig}\xspace}
\newcommand{\comphep}   {\textsc{comphep}\xspace}

\newcommand{\wplus}     {\ensuremath{W+}jets\xspace}

\def\msb{{\bar{\ssstyle M \kern -1pt S}}}

\begin{document}
\begin{titlepage}
\pubblock

\vfill
\Title{Measurement of top quark polarization in ttbar lepton+jets final states at D0}
\vfill
\Author{ Kamil Augsten \\ for the D0 Collaboration}
\Address{\institute}
\vfill
\begin{Abstract}
We present a measurement of top quark polarization in \ttbar pair production in \ppbar collisions at $\sqrt{s}=1.96$\,TeV using data corresponding to 9.7\,fb$^{-1}$ of integrated luminosity recorded with the D0 detector at the Fermilab Tevatron Collider. We consider final states containing a lepton and at least three jets. The polarization is measured through the distribution of lepton angles along three axes: the beam axis, the helicity axis, and the transverse axis normal to the \ttbar production plane. This is the first measurement of top quark polarization at Tevatron using lepton+jet final states and the first measurement of the transverse polarization in $t\overline t$ production. The observed distributions are consistent with standard model predictions of nearly no polarization. 
\end{Abstract}
\vfill
\begin{Presented}
$9^{th}$ International Workshop on Top Quark Physics\\
Olomouc, Czech Republic,  September 19--23, 2016
\end{Presented}
\vfill
\end{titlepage}
\def\thefootnote{\fnsymbol{footnote}}
\setcounter{footnote}{0}

The standard model (SM) predicts that top quarks produced at the Tevatron collider are almost unpolarized,
while models beyond the standard model (BSM) predict enhanced polarizations~\cite{Fajfer:2012si}.
The top quark polarization $P_{\hat{n}}$ can be measured in the top quark rest frame through the angular distributions of the top quark decay products relative to some chosen axis $\hat{n}$ \cite{Bernreuther}, $\frac{1}{\Gamma}\frac{d\Gamma}{d\cos{\theta_{i,\hat{n}}}}=\frac{1}{2}(1+P_{\hat{n}}\kappa_{i}\cos{\theta_{i,\hat{n}}})$,
where $i$ is the decay product (lepton, quark, or neutrino), $\kappa_{i}$ is its spin analyzing power ($\approx 1$ for charged leptons, 0.97 for $d$-type quarks, $-0.4$ for $b$-quarks, and $-0.3$ for neutrinos and $u$-type quarks \cite{Brandenburg}), and $\theta_{i,\hat{n}}$ is the angle between the direction of the decay product $i$ and the quantization axis $\hat{n}$. 

We measure the polarization in angular distribution of leptons along three quantization axes: \textrm{(i)} {\bf beam axis $\hat{n}_{p}$}, given by the direction of the proton beam~\cite{Bernreuther}, \textrm{(ii)} {\bf helicity axis $\hat{n}_{h}$}, given by the direction of the parent top or antitop quark, and \textrm{(iii)} {\bf transverse axis $\hat{n}_T$}, given as perpendicular to the production plane defined by the proton and parent top quark directions, \ie ,
$\hat{n}_{p} \times \hat{n}_{t}$ (or by $\hat{n}_{p} \times - \hat{n}_{\overline{t}}$ for the antitop quark)~\cite{Bernreuther:1995cx}.

The longitudinal polarizations along the beam and helicity axes at the Tevatron collider are predicted by the SM to be $(-0.19 \pm 0.05) \%$ and $(-0.39 \pm 0.04) \%$~\cite{weakcorr}, respectively, while the transverse polarization is estimated to be $\approx 1.1 \%$~\cite{Baumgart:2013yra}.
The polarization at the Tevatron and LHC are expected to be different because of the difference in the initial states, which motivates the measurement of the polarizations in Tevatron data~\cite{top2014}. For beam and transverse axes, the top quark polarizations in \ppbar collisions are expected to be larger than those for $pp$~\cite{Bernreuther,Bernreuther:1995cx}, therefore offering greater sensitivity to BSM models.

Each top quark of the \ttbar pair decays into a $b$ quark and a $W$ boson with nearly $100\%$ probability. In $\ell$+jets events, one of the $W$ bosons decays leptonically and the other into quarks that evolve into jets. 
This analysis requires the presence of one isolated $e$ or $\mu$ with  transverse momentum $\pt>20$\,GeV and physics pseudorapidity $|\eta|<1.1$ or $|\eta|<2$, respectively. 
We require at least three jets with $\pt>20$\,GeV within $|\eta|<2.5$, and $\pt>40$\,GeV for the jet of highest \pt. At least one jet per event is required to be identified as originating from a $b$ quark ($b$ tagged). An imbalance in transverse momentum $\met>20$\,GeV is expected from the undetected neutrino. Additional quality requirements are applied. The detailed description of all requirements can be found in Refs.~\cite{bib:diffxsec,Polar}.

We simulate \ttbar events at the NLO with the \mcatnlo event generator version 3.4 \cite{mcatnlo} with \herwig \cite{herwig} for parton showering, hadronization, and modeling of the underlying event. The background processes are generated with \alpgen \cite{alpgen}, \pythia  \cite{pythia}, and \comphep \cite{comphep}, or estimated from the data  in case of multijet background.
Six different BSM models~\cite{Axi} are used to study modified \ttbar production: one $Z'$ boson model and five axigluon models with different axigluon masses and couplings (m200R, m200L, m200A, m2000R, and m2000A). 

A constrained kinematic $\chi ^{2}$ fit is used to associate the observed leptons and jets with the individual top quarks using a likelihood term for each jet-to-quark assignment. 
The algorithm includes a technique that reconstructs events with a lepton and only three jets~\cite{rocfit}. All possible assignments of jets to final state quarks are considered and weighted by the $\chi ^{2}$ probability of each kinematic fit.

To determine the sample composition, we construct a kinematic discriminant based on the approximate likelihood ratio of expectations for \ttbar and \wplus events. The input variables are chosen to have good separation power, to be well modeled and not strongly correlated. The kinematic variables and details about the method are described in \cite{Polar}.
The sample composition is determined from a simultaneous maximum-likelihood fit to the discriminant distributions.
The \wplus background is normalized separately for the heavy-flavor and light-parton contributions. The sample composition after implementing the selections is summarized in Table~\ref{Tab:Sample}. The obtained \ttbar yield is close to the expectations.

\begin{wraptable}{r}{11cm}
\begin{tabular}{lcccc}
          & \multicolumn{2}{c}{3 jets} & \multicolumn{2}{c}{$\geq 4$ jets} \\
Source    & $e$+jets & $\mu$+jets & $e$+jets & $\mu$+jets \\ \hline
$W$+jets          &  $1741 \pm 26$       & $1567 \pm 15$         &  $339 \pm 3$       &  $295 \pm 3$        \\
Multijet        &  $494 \pm 7$        & $128 \pm 3$          &  $147 \pm 4$         &  $49 \pm 2$        \\
Other Bkg       &  $446 \pm 5$       & $378 \pm 2$          &  $87 \pm 1$        &  $73 \pm 1$        \\
$t\overline{t}$ signal &  $1200 \pm 25$       & $817 \pm 20$         &  $1137 \pm 24$      &  $904 \pm 23$        \\ \hline
Sum      &  $3881 \pm 37$      & $2890 \pm 25$        &  $1710 \pm 25$      &  $1321 \pm 23 $        \\ \hline
Data            &  $3872$       &  $2901$        &  $1719$       &    $1352$      \\
\end{tabular}
\caption{
 Sample composition and event yields with statistical uncertainties after implementing the selection requirements and the maximum likelihood fit.
}
\label{Tab:Sample}
\end{wraptable}

The lepton angular distributions in leading background, the \wplus events, are reweighted so that the $\cos{\theta_{\ell,\hat{n}}}$ distributions agree with those for the control events in data with \ttbar and other background components subtracted. The control events are $\ell$+3 jet without $b$ tagged jets that are dominated by \wplus process with $>70\%$ contribution.

To measure the polarization, a fit is performed to the reconstructed $\cos{\theta_{\ell,\hat{n}}}$ distribution using \ttbar templates of $+1$ and $-1$ polarizations, and background templates normalized to the expected yields. The MC signal templates, generated with no polarization, are reweighted to the expected double differential distribution~\cite{Bernreuther}:
\begin{equation}
\frac{1}{\Gamma} \frac{d\Gamma}{d\cos{\theta_{1}}\cos{\theta_{2}}} = \frac{1}{4}(1+\kappa_{1}P_{\hat{n},1}\cos{\theta_{1}}+ \\ +\rho\kappa_{2}P_{\hat{n},2}\cos{\theta_{2}}-\kappa_{1}\kappa_{2}C\cos{\theta_{1}}\cos{\theta_{2}}),
\label{Eq.RW}
\end{equation}
where indices 1 and 2 represent the $t$ and $\bar{t}$ quark decay products and $C$ is the SM \ttbar spin correlation coefficient for a given quantization axis (helicity $C=-0.368$, beam $C=0.791$, transverse not know. thus $C=0$ and systematic uncertainty based on this choice). The $P_{\hat{n},i}$ represents the polarization state we model, $\pm 1$. 
In the SM, assuming $CP$ invariance, $P_{\hat{n},1} =  P_{\hat{n},2}$ and gives the relative sign factor $\rho$ a value of +1 for the helicity axis and $-1$ for the beam and transverse axes.

\begin{figure}[!ht]
\begin{centering}
\includegraphics[width=0.32\columnwidth]{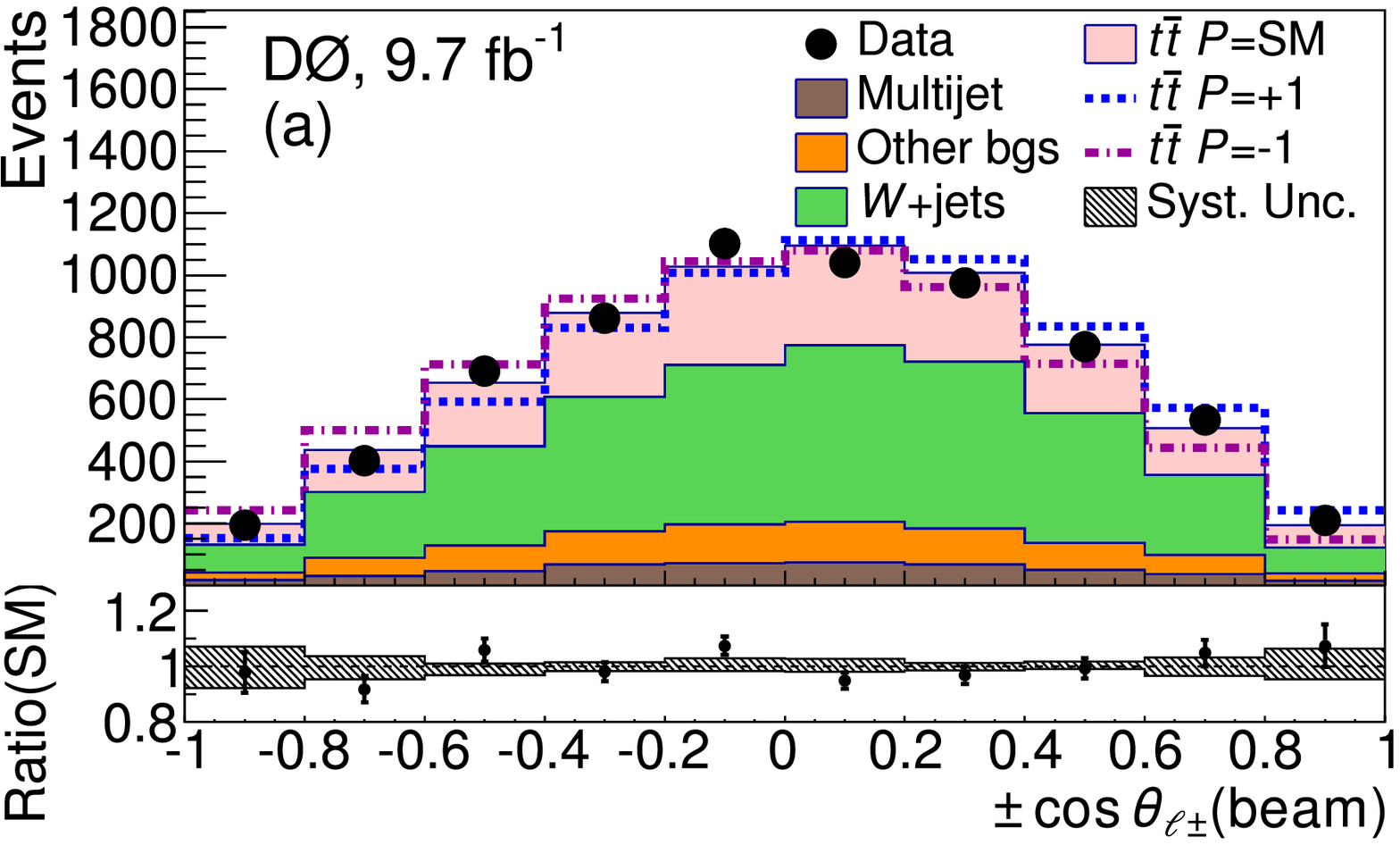}
\includegraphics[width=0.32\columnwidth]{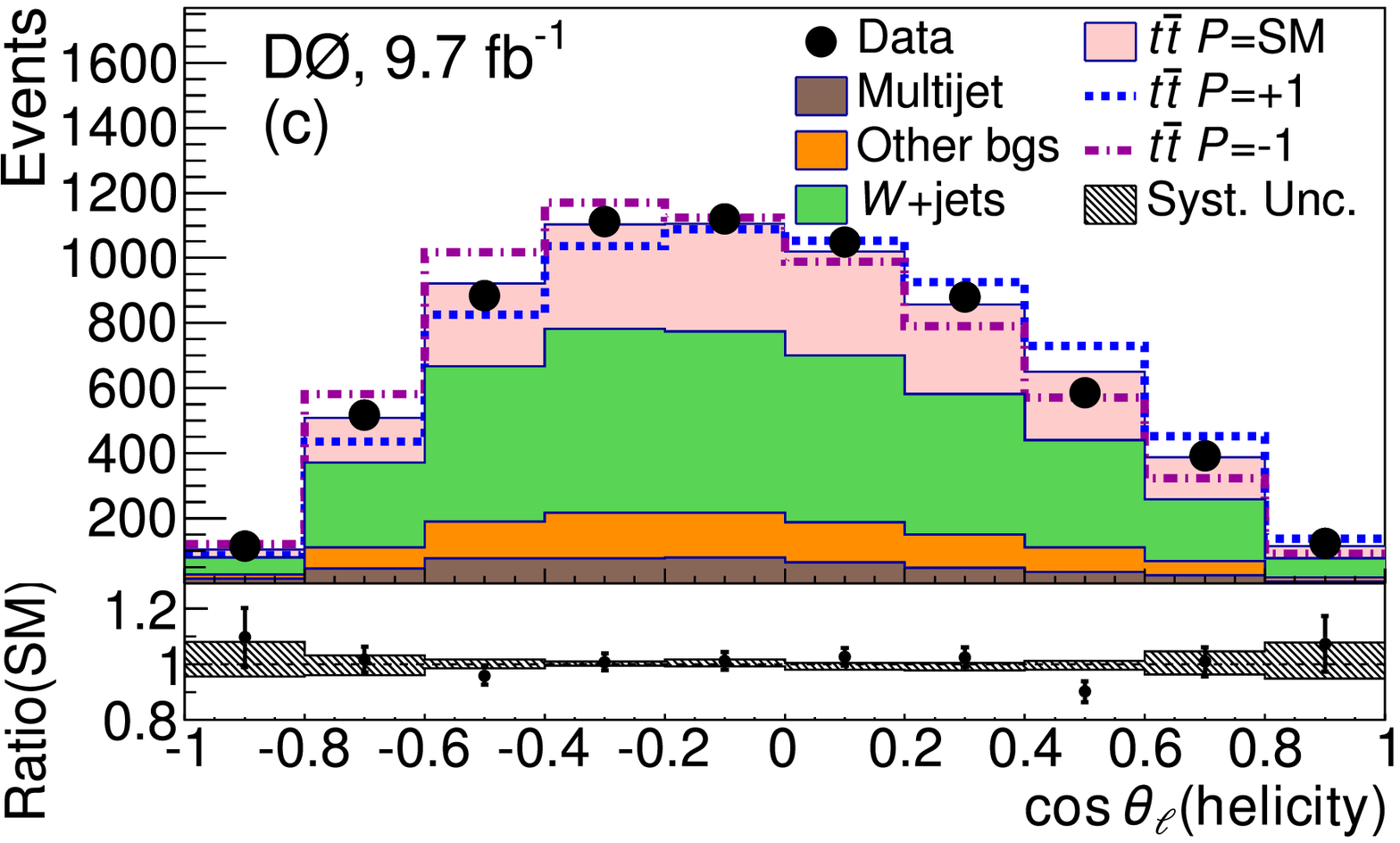}
\includegraphics[width=0.32\columnwidth]{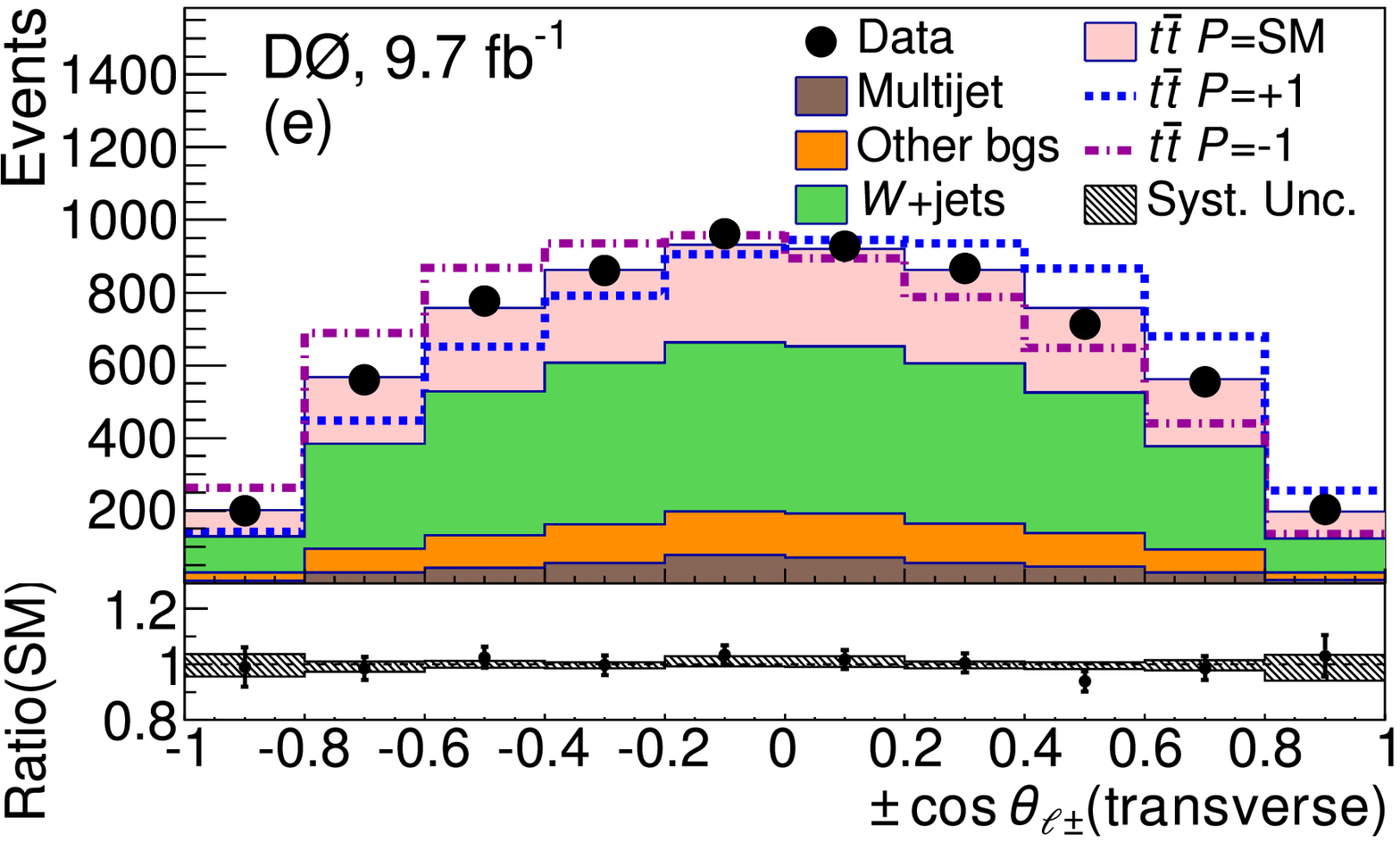}\\
\includegraphics[width=0.32\columnwidth]{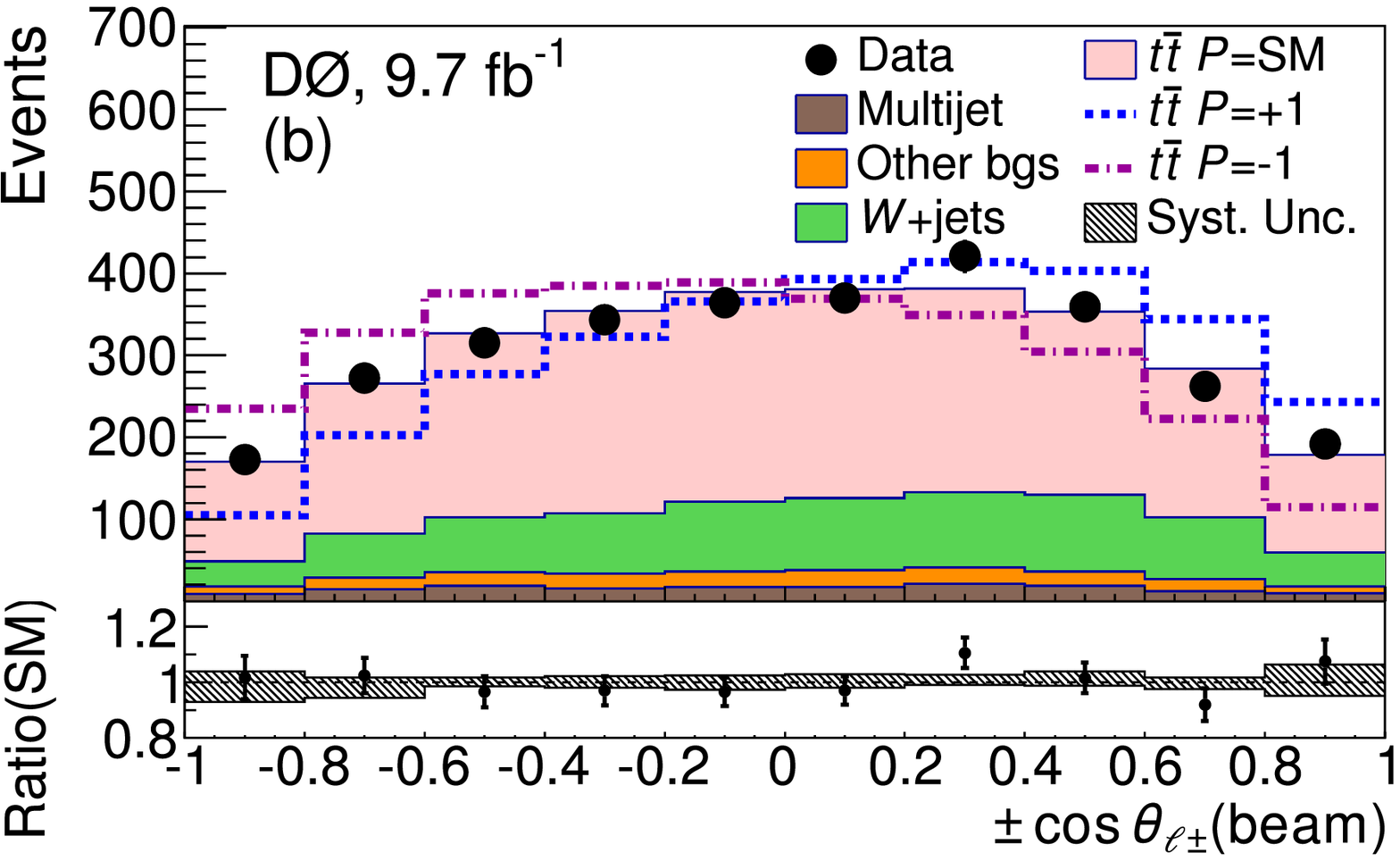} 
\includegraphics[width=0.32\columnwidth]{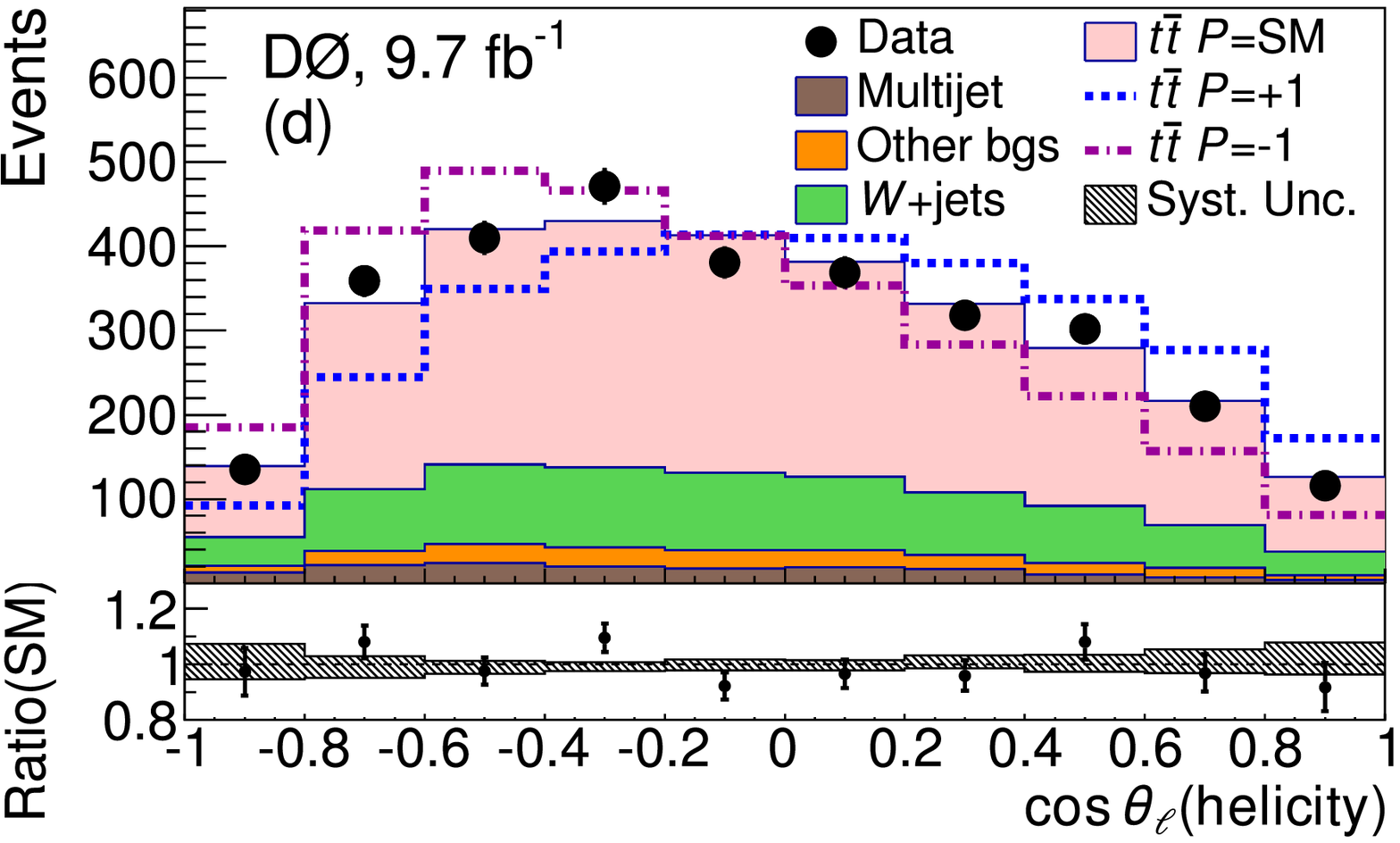} 
\includegraphics[width=0.32\columnwidth]{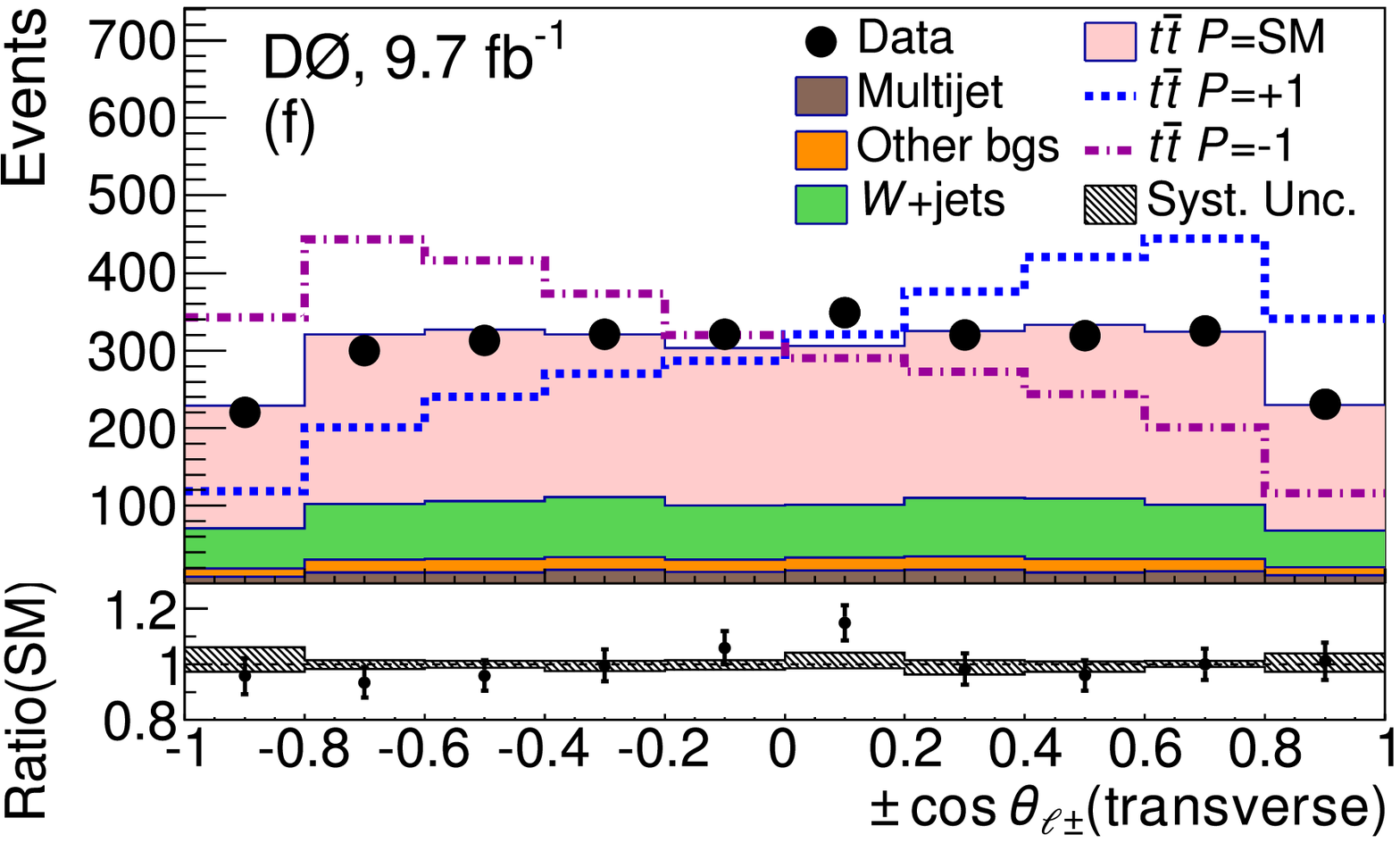}\\
\par\end{centering}
\caption{
\label{fig:templatefit}
The $\ell$+jets $\cos\theta$ distributions for data, expected backgrounds, and signal templates $P=-1$, SM, and +1.
(a), (c), and (e) show $\ell$+3 jet events; (b), (d), and (f) show $\ell + \geq 4$ jet event.
The hashed areas represent systematic uncertainties. The direction of the $\cos\theta$ axis is reversed for the $\ell^-$ events for beam and transverse axes.
}
\end{figure}

A simultaneous fit is performed for the eight samples defined according to lepton flavor, lepton charge, and number of jets. The observed polarization is taken as $P=f_+ - f_-$, where $f_\pm$ are the fraction of $P=+1$ and $-1$ events returned from the fit. The fitting procedure and method are verified using pseudo-experiments for five values of polarization, and through a check of consistency with predictions, using the BSM models with non-zero generated longitudinal polarizations.
The distributions in the cosine of the polar angle of leptons from \ttbar decay for all three axes are shown in Fig.~\ref{fig:templatefit}.

\begin{table}[!ht]
\begin{centering}
\begin{tabular}{lccc}
Source & Beam & Helicity & Transverse \\
\hline
 ~~Jet reconstruction			& $\pm0.010$ & $\pm0.008$ & $\pm0.008$ \\
 ~~Jet energy measurement		& $\pm0.010$ & $\pm0.023$ & $\pm0.006$ \\
 ~~$b$ tagging				& $\pm0.009$ & $\pm0.014$ & $\pm0.005$ \\
 ~~Background modeling			& $\pm0.007$ & $\pm0.021$ & $\pm0.004$ \\
 ~~Signal modeling			& $\pm0.016$ & $\pm0.020$ & $\pm0.008$ \\ 
 ~~PDFs					& $\pm0.013$ & $\pm0.011$ & $\pm0.003$ \\
 ~~Methodology				& $\pm0.013$ & $\pm0.007$ & $\pm0.009$ \\
\hline 
~~Total systematic uncertainty 		& $\pm0.030$ & $\pm0.042$ & $\pm0.017$ \\
~~Statistical uncertainty		& $\pm0.046$ & $\pm0.044$ & $\pm0.030$ \\
~~Total uncertainty			& $\pm0.055$ & $\pm0.061$ & $\pm0.035$ \\

\end{tabular}
\par\end{centering}
\caption{
\label{tab:syst}
Summary of the top quark polarization measurement uncertainties. Groups of systematic sources and the total uncertainties are shown.
}
\end{table}

\begin{wrapfigure}{l}{0.5\textwidth}
 \includegraphics[width=0.95\linewidth]{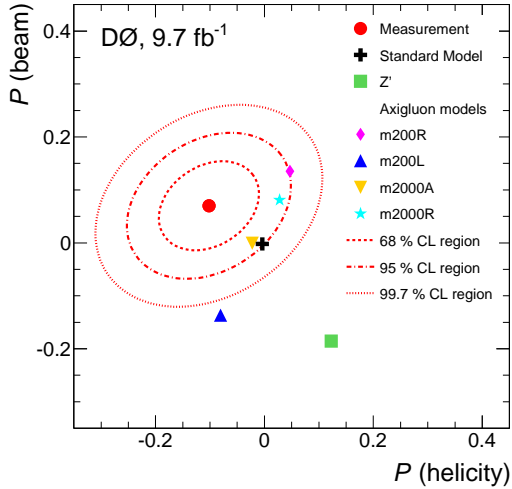}
 \caption{
 \label{fig:final}
Two dimensional visualization of the longitudinal top quark polarizations in the $\ell$+jets channel along the beam and helicity axes compared with the SM and the BSM models. The correlation of the two measurement uncertainties is $27 \%$.
 }
\end{wrapfigure}

We perform correction for the difference between the nominal \mcatnlo forward-backward $t$ and $\bar{t}$ asymmetry $A_{\mathrm{FB}}$ of $(5.01 \pm 0.03) \%$ and the NNLO calculation \cite{AfbNNLO} of $(9.5 \pm 0.7) \%$ as correlation between top quark polarization and $A_{\mathrm{FB}}$ has been observed \cite{Boris}.

Several categories of systematic uncertainties have been evaluated using fully simulated events, listed in Table~\ref{tab:syst}. Details about the evaluation of the uncertainties can be found in Ref.~\cite{bib:diffxsec, Polar}.

The measured polarizations for the three spin quantization axes are shown in Table~\ref{Tab:final} together with combination with the previous D0 result in the dilepton channel~\cite{Boris} for the beam axis. Results on the longitudinal polarizations are presented in Fig.~\ref{fig:final}. The polarizations are consistent with SM predictions. The transverse polarization was measured for the first time. These are the most precise measurements of top quark polarization in \ppbar collisions.

\begin{table}[!ht]
\begin{center}
\begin{tabular}{l c c}
Axis    & Measured polarization & SM prediction \\ \hline
Beam             & $+0.070 \pm 0.055$ & $-0.002$    \\
\,\textit{Beam - D0 comb.}             & $+0.081 \pm 0.048$ & $-0.002$    \\
Helicity             & $-0.102 \pm 0.061$ & $-0.004$    \\
Transverse             & $+0.040 \pm 0.035$ & $+0.011$    \\
\end{tabular}
\end{center}
\caption{
  Measured top quark polarization from the \ttbar $\ell$+jet channel along the beam, helicity, and transverse axes, and the combined polarization for beam axis with the dilepton result by the D0 Collaboration~\cite{Boris}. 
}
\label{Tab:final}
\end{table}


\end{document}